\documentclass{article}

\usepackage{amstext}
\usepackage{amsmath}
\usepackage{amssymb}
\usepackage{amsthm}

\date{}

\title{Computing Clebsch-Gordan matrices with applications in elasticity theory}
\author{S. Selivanova\thanks{This
research was partially supported by the Federal Target Grant
``Scientific and educational personnel of innovative Russia'' for
2009-2013, program 1.2.2, contract No. 8217, and by the Russian Foundation for Basic Research, project No. 12-01-31183.}}

\begin{document}

\maketitle

\begin{abstract}
We provide an algorithm of computing Clebsch-Gordan
coefficients for irreducible representations, with integer
weights, of the rotation group $SO(3)$ and demonstrate the convenience of this algorithm for constructing new (to our knowledge) models in anisotropic elasticity theory.

\end{abstract}

\section{Introduction }

The aim of this paper is to provide an algorithm of
computing Clebsch-Gordan coefficients, written in a convenient
real-valued matrix form, for irreducible representations, with
integer weights, of the rotation group $SO(3)$, and to show its convenience for modelling anisotropic elasticity problems. 

The Clebsch-Gordan coefficients, arising from Kronecker products of
representations (see Section 2 for definitions), are a classical notion in group
representation theory, and they are widely used in mathematical
physics, see e.g. \cite{ed,gelf,shel1,mil,shel, stern, vil}.
 In  \cite{gg, gm, gmr, mikh} it is shown how, written in a matrix form,
they allow to construct invariant differential
operators (as $\frac{\partial}{\partial x_i}G^i_{1[L,L\pm1]}$, where $G^i_{1[L,L\pm1]}$ are particular examples of Clebsch-Gordan matrices, see definitions and examples in the paper) and matrix spherical functions (as products of relevant Clebsch-Gordan matrices and homogeneous polynomials). It helps to write many (systems of) first-order PDEs invariant under rotations in a convenient form and to find explicit solutions of Cauchy problems for them, see e.g.  discussions on the examples  of elasticity, acoustics and Maxwell equations in the mentioned  references.

The algorithm  described in Section 4 relies on formulas derived recently in the papers \cite{gord2,gord1} (which were written in continuation of the monograph \cite{gm}), where a special basis, w.r.t.  which the matrices of
the representations are real orthogonal (in contrast to complex
unitary matrices in most literature on mathematical physics), is proposed.
The construction of this basis and further technical details are described in Sections 2, 3. The general idea of the algorithm  consists of calculating highest-/lowest-weight coefficients first and then applying lowering/raising operators and is well-known (see e.g. \cite{ed, gelf,vil} and modern improvements \cite{al,gli}). The (probably) new aspect is the special basis of \cite{gord2,gord1} and the spinor technique we use for our computations. In comparison to other recent papers on the subject, the presentation is very elementary, and possibility of application to (linear and nonlinear) elasticity theory is demonstrated (Section 5).

In particular, we use the Clebsch-Gordan  matrices to construct quadratic
invariants and write down an equation of state of an anisotropic
elastic media, as a thermodynamical potential depending on these
invariants. By using purely group representation methods we are able to write down the precise form of the invariants for all
crystal systems, without any complicated geometric considerations.
The parameters of the invariant quadratic forms may be interpreted
as the parameters of the media from the Hook's law (Lame
coefficients in the isotropic case), in accordance with well-known facts of elasticity theory \cite{sed,lan, steig}. 

This construction can be
used to write elasticity equations in an invariant form which is
convenient for modelling complicated physical processes like exploding welding of metals and for applying numerical methods (the work in this direction is in progress and will apear in forthcoming publications). These considerations are a development of the paper \cite{ss08} and are in line with the works \cite{god,gp,gr,p}.

\section{Preliminaries}

Let us first recall some basic notions following the book
\cite{gm}. The {\it rotations} of a three-dimensional Euclidean
space $\mathbb R^3$ constitute the group
$$SO(3)=\{g\in GL(3)\ |\ g^Tg=I_3,\ \det(g)=1\},$$ where $GL(3)$ is the
group of all nondegenerate $3\times 3$ matrices, while $I_3$ is
the unit matrix.

 It is said that we have a  {\it representation} $T_G$ of a group
$G$ in a $k$-dimensional vector space ${\cal L}$, if to each element $g\in G$
 there corresponds a linear mapping $T_g:{\cal
L}\to{\cal L}$ such that
\begin{equation}\label{T}T_{e}=I_k,\ \ T_{g_1\cdot g_2}=T_{g_1}\cdot T_{g_2},\end{equation}
where $e$ is the identity of the group $G$.

 The representation $T_G$ is called  {\it
irreducible}, if in ${\cal L}$ there are no nontrivial subspaces
invariant under all mappings $T_g$, where $g\in G$. The number $N$
(integer or half of an integer), such that  $k=2N+1$, is called
the {\it weight} of the irreducible representation. It is
well-known that
 single-valued irreducible representations of
$SO(3)$ can exist only in odd-dimensional spaces (i.e. when $N$ is
integer), moreover, for a fixed number $N$, the corresponding
representation is unique up to equivalence. In this paper we
consider only integer weights $N$.

The elements $\{g(t)\in G\}_{t\in\mathbb R}$ constitute a {\it
one-parameter subgroup} of the group $G,$ if 
\begin{equation}\label{g}g(t+s)=g(t)\cdot
g(s)\text{ and }g(0)=e.\end{equation}
Let ${\cal L}_N$ be the ($2N+1$)-dimensional vector space in which
the action of an irreducible representation $T_G$ of $G=SO(3)$ is
defined, and let us denote the Cartesian coordinates in $\mathbb
R^3$ as $x_{-1},x_0,x_1.$ Then the one-parameter subgroups
corresponding to rotations w.r.t. these axes are described by the
families of matrices
$$g_{1}(\omega)=\left[\begin{array}{ccc}
1&0&0\\ 0&cos(\omega)&-sin(\omega)\\ 0&sin(\omega)&cos(\omega)
\end{array}\right], \
g_{-1}(\omega)=\left[\begin{array}{ccc}
cos(\omega)&0&sin(\omega)\\0&1&0\\  -sin(\omega)&0&cos(\omega)
\end{array}\right], $$
$$g_{0}(\omega)=\left[\begin{array}{ccc}
 cos(\omega)&-sin(\omega)&0\\ sin(\omega)&cos(\omega)&0\\0&0&1
\end{array}\right],$$
where $\omega\in\mathbb R$.

Denote as $J_j$, $j=-1,0,1$, the  {\it infinitesimal operators} of
the representations of these subgroups:
\begin{equation}\label{inf_oper}J_j=\frac{d}{d\omega}T_{g_{j}(\omega)}|_{\omega=0}.\end{equation}

There are different possibilities to chose a canonical, w.r.t. the
representation $T_G$, basis in ${\cal L}_N$. A standard way, see
e.g. \cite{gelf,gm,vil}, is to chose vectors on which the
infinitesimal operators act as follows

\begin{equation}\label{e_C}iJ_0\,e_N^n\,=\ ne_N^n,\
n=-N,\ldots,N,\end{equation}
\begin{equation}\label{e_down}(iJ_{1}+J_{-1})\,e_N^n\,=\,
\begin{cases}-\sqrt{(N-n+1)(N+n)}\,e_N^{n-1},\
\ n=-N+1,\ldots,N,\\
0,\ \ n=-N
\end{cases}\end{equation}
\begin{equation}\label{e_up}(iJ_1-J_{-1})\,e_N^n\,=\,
\begin{cases}-\sqrt{(N-n)(N+n+1)}\,e_N^{n+1},\
\ n=-N,\ldots,N-1,\\
0,\ \ n=N.
\end{cases}\end{equation}

Note that $e^n_{N}$  are eigenvectors of the operator $J_0$ (they
are defined up to a multiplicative constant), while formulas
\eqref{e_down} and \eqref{e_up}  allow to compute the  basis
vectors recursively. W. r. t. such basis, which is called {\it a
canonical basis of ``e'' type}, $T_G$ is defined by unitary
matrices of dimension $(2N+1)\times (2N+1)$.

In \cite{gord2} a basis, w.r.t. to which $T_G$ is defined by
(real) orthogonal matrices, is introduced, and it is called a {\it
canonical basis of ``h'' type}. This basis is related to the
previous one by means of a unitary transformation:
\begin{equation}\label{h}\begin{cases}h_N^{-n}=\frac{(-i)^{N-1}}
{\sqrt{2}}\left[(-1)^ne_N^n-e_N^{-n}\right],\\
h_N^0=(-i)^Ne_N^0,\\
h_N^n=\frac{-(-i)^{N}}{\sqrt{2}}\left[(-1)^ne_N^n+
e_N^{-n}\right].\end{cases}\end{equation}

The action of the infinitesimal operators on this basis is as
follows:
\begin{equation}\label{h_C}
\left(\begin{array}{cc}0&J_0\\-J_0&0 \end{array}\right)
\left(\begin{array}{c}h_N^{-n}\\h_N^n\end{array}\right)=
n\left(\begin{array}{c}h_N^{-n}\\h_N^n\end{array}\right), n\geq0;
\end{equation}
\begin{equation}\label{h_down}
\begin{cases}
\left(\begin{array}{cc}J_{-1}&J_1\\-J_1&J_{-1} \end{array}\right)
\left(\begin{array}{c}h_N^{-n}\\h_N^n\end{array}\right)=
\sqrt{(N+n)(N-n+1)}\left(\begin{array}{c}h_N^{-n+1}\\h_N^{n-1}\end{array}\right),
n\geq2, \vspace{6.pt}\\
\left(\begin{array}{cc}J_{-1}&J_1\\-J_1&J_{-1} \end{array}\right)
\left(\begin{array}{c}h_N^{-1}\\h_N^1\end{array}\right)=
-\sqrt{2N(N+1)}\left(\begin{array}{c}0\\ h_N^0\end{array}\right),\
\end{cases}
\end{equation}
\begin{equation}\label{h_up}
\begin{cases}
\left(\begin{array}{cc}J_{-1}&-J_1\\J_1&J_{-1} \end{array}\right)
\left(\begin{array}{c}h_N^{-n}\\h_N^n\end{array}\right)=
-\sqrt{(N+n+1)(N-n)}\left(\begin{array}{c}h_N^{-n-1}\\h_N^{n+1}\end{array}\right),
n\geq1, \vspace{6.pt}\\
\left(\begin{array}{cc}J_{-1}&J_1\\J_1&J_{-1} \end{array}\right)
\left(\begin{array}{c}0\\ h_N^0\end{array}\right)=
\sqrt{\frac{N(N+1)}{2}}\left(\begin{array}{c}h_N^{-1}\\h_N^1\end{array}\right),\
\end{cases}
\end{equation}

Recall that a {\it Kronecker product} $T_g=T_g^1\times T_g^2$ of
the representation $T_g^1$ of weight $N_1$ and $T_g^2$ of weight
$N_2$ is a representation which acts on a matrix $B$ of dimension
$(2N_1+1)\times(2N_2+1)$ as
\begin{equation}\label{TB}T_g\,B=T_g^1\,B\,(T_g^2)^T.\end{equation}
The definition of the Clebsch-Gordan matrices, that we are going to
compute, arises from the following statement \cite{gord2}.

 {\bf Theorem.} {\sl If the representations $T_g^1$
and $T_g^2$ are irreducible, then $T_g^1\times T_g^2$ can be
decomposed into a direct sum of irreducible representations of the
following weights:}
\begin{equation}\label{vesa}N=|N_1-N_2|,\
|N_1-N_2|+1,\ldots,N_1+N_2.\end{equation} Such decomposition is
realized by means of the {\it Clebsch-Gordan matrices}, which
constitute canonical bases of the corresponding spaces of
matrices:
\begin{equation}G_{N[N_1,N_2]}^n\text{ of dimension }(2N_1+1)\times(2N_2+1),\
n=-N,N+1,\ldots,N.\end{equation} These matrices are real-valued, orthonormal:
\begin{equation}tr\left\{(G_{N[N_1,N_2]}^n)^TG_{N[N_1,N_2]}^m\right\}=
\delta_{mn}\ \text{(the Kronecker symbol)}, \end{equation} possess one or two
non-zero diagonals, and satisfy the following symmetry property:
\begin{equation}\label{k_g}G_{N[N_1,N_2]}^n=(-1)^{N+N_1+N_2}(G_{N[N_2,N_1]}^n)^T.
\end{equation}
For computations, different realizations of the irreducible
representations are useful, in particular: in the space of
homogeneous polynomials of three real variables; in the
space of homogeneous spinor polynomials of two complex variables;
in the space of matrices. Our computations rely on the last two
ones; we describe them in the next section.

\section{Different realizations of irreducible representations of the rotation group}

A {\it spinor polynomial} is a homogeneous polynomial
$f(\xi,\eta)$ of degree $2N$ of two complex variables $\xi,\eta$.
The dimension of this space is equal to $2N+1$.

As it is known, the elements of $SO(3)$ can be parameterized by
unitary matrices $g\in SU(2)$ in such a way that to any
irreducible representation of $SU(2)$ of an integer weight $N$
there corresponds a unique representation of $SO(3)$ in a space of
odd dimension equal to $2N+1$. This parametrization helps to
describe all single-valued irreducible representations of $SO(3)$,
since it is sometimes more convenient to make computations for
$SU(2)$. For simplicity of notation we denote the one-parameter
subgroups of $SU(2)$ and the corresponding infinitesimal operators
in the same way $g_j(w)$ and $J_j$, as for $SO(3)$ above.

In the space of spinor polynomials the irreducible representation
of $SU(2)$ of weight  $N$ is realized by the following formula
\cite{gord2}: for 
\begin{equation}g\in SU(2),\
g=\large\left(\begin{array}{cc}\alpha&\beta\\
-\overline{\beta}&\overline{\alpha}\end{array}\right)\end{equation} we have
\begin{equation}T_gf(\xi,\eta)=f(\xi^{\ '},\eta^{\ '})\equiv f(\overline{\alpha}
\xi-\beta\eta,\overline{\beta}\xi+\alpha\eta).\end{equation} The canonical
basis of ``e'' type consists of the polynomials
\begin{equation}\label{e}e_N^n(\xi,\eta)=\tilde{\rho}(N,n)\xi^{N+n}
\eta^{N-n}=(-1)^{N+n}
\sqrt{\frac{(2N+1)!}{(N+n)!(N-n)!}}\xi^{N+n}\eta^{N-n},\end{equation}
where $n=-N,\ldots N$. According to \eqref{h}, the canonical basis
of ``h'' type is defined as
\begin{equation}\label{h_e}h_N^{\cdot}=e_N^{\cdot}U_N,\end{equation}
\begin{equation}\label{e_h}e_N^{\cdot}=h_N^{\cdot}V_N,\end{equation}
where
$$h_N^{\cdot}=\left(h_N^{-N},h_N^{-N+1},\ldots,h_N^0,\ldots,h_N^N\right),\
e_N^{\cdot}=\left(e_N^{-N},e_N^{-N+1},\ldots,e_N^0,\ldots,e_N^N\right),$$
 $$U_N=\small\frac{(-i)^{N-1}}{\sqrt{2}}\left[
\begin{array}{ccccccc}
-1    &      &      &          &       &      &i      \\
      &\ldots&      &          &       &\ldots&       \\
      &      &-1    &          &i      &      &       \\
      &      &      &-i\sqrt{2}&       &      &       \\
      &      &(-1)^n&          &i(-1)^n&      &       \\
      &\ldots&      &          &       &\ldots&       \\
(-1)^N&      &      &          &       &      &i(-1)^N\
\end{array}\right],$$
$ U_N^*U_N=I_{2N+1}$ and
$$V_N=U_N^{-1}=U_N^*=
\small\frac{i^{N-1}}{\sqrt{2}}\left[
\begin{array}{ccccccc}
-1    &      &      &          &        &      &(-1)^N  \\
      &\ldots&      &          &        &\ldots&        \\
      &      &-1    &          &(-1)^n  &      &        \\
      &      &      & i\sqrt{2}&        &      &        \\
      &      &    -i&          &-i(-1)^n&      &        \\
      &\ldots&      &          &        &\ldots&        \\
-i    &      &      &          &        &      &-i(-1)^N
\end{array}\right].$$

The product of two irreducible representations of weights $N_1$
and $N_2$ (of dimensions $2N_1+1$ and $2N_2+1$), can be realized
in the space of {\it bispinor} polynomials
$f(\xi_1,\eta_1;\xi_2,\eta_2),$ homogeneous of degrees $2N_1$ and
$2N_2$ on $\xi_1,\eta_1$ and $\xi_2,\eta_2$, respectively:
\begin{equation}T_gf(\xi_1,\eta_1;\xi_2,\eta_2)=f(\overline{\alpha}\xi_1-\beta\eta_1,
\overline{\beta}\xi_1+\alpha\eta_1;
\overline{\alpha}\xi_2-\beta\eta_2,\overline{\beta}\xi_2+\alpha\eta_2).\end{equation}This representation acting in a
$(2N_1+1)\cdot(2N_2+1)$-dimensional space is reducible and it can
be decomposed into irreducible ones as in \eqref{vesa}. The
canonical bases of ``e'' type look as \cite{gm,gg}
\begin{equation}\label{mu}
\mu_{N[N_1,N_2]}^n(\xi_1,\eta_1;\xi_2,\eta_2)=
\end{equation}
$$=\hat{\rho}(N,N_1,N_2)(\xi_1\eta_2-\eta_1\xi_2)^{N_1+N_2-N}
\left(-\xi_2\frac{\partial}{\partial\xi_1}-\eta_2\frac{\partial}
{\partial\eta_1}\right)^{N-N_1+N_2}e_N^n(\xi_1,\eta_1),$$ where
$n=-N,-N+1,\ldots,N,$ and $N_1,N_2$ are fixed and $N$ is as in
\eqref{vesa};
$$\hat{\rho}(N,N_1,N_2)=\sqrt{\frac{(2N_1+1)!(2N_2+1)!(N+N_1-N_2)!}
{2N!(N_1+N_2-N)!(N-N_1+N_2)!(N+N_1+N_2+1)!}}.$$ 

The complete proof of this fact can be found in \cite{gm} (Section 13). Since this book is in Russian, we recall briefly its main steps.

$\bullet$ First, the infinitesimal operators acting in the space of bispinor polynomials are calculated (for spinor polynomials it is a straightforward calculation by definition, and for the bispinor case the definition of the Kronecker product of representations is used). The answer is as follows:
\begin{equation}(iJ_{1}+J_{-1})=\eta_1\frac{\partial}{\partial \xi_1}+\eta_2\frac{\partial}{\partial \xi_2},\quad (iJ_{1}-J_{-1})=\xi_1\frac{\partial}{\partial \eta_1}+\xi_2\frac{\partial}{\partial \eta_2},\end{equation}
\begin{equation}iJ_0=\frac{1}{2}\left(\xi_1\frac{\partial}{\partial \xi_1}-\eta_1\frac{\partial}{\partial \eta_1}\right)+\frac{1}{2}\left(\xi_2\frac{\partial}{\partial \xi_2}-\eta_2\frac{\partial}{\partial \eta_2}\right).\end{equation}

$\bullet$ It is noticed (by direct computation of the action of the infinitesimal operators) that the products $e_{N_1}^{n_1}e_{N_2}^{n_2}$ do not constitute a canonical basis in the space of bispinor polynomials. The form of the polynomials needed is ``guessed'' by taking in account the identities
\begin{equation}(iJ_{1}+J_{-1})(\xi_1\eta_2-\xi_2\eta_1)^q=0,\quad 
(iJ_{1}-J_{-1})(\xi_1\eta_2-\xi_2\eta_1)^q=0\end{equation}
and by choosing a proper norming after calculating the scalar square of the polynomial
\begin{equation}(\xi_q\eta_2-\xi_2\eta_1)^a\eta_1^b\eta_2^c.\end{equation}

$\bullet$ Finally, a direct substitution of the basis \eqref{mu} into the equalities \eqref{e_C}--\eqref{e_down} shows that it satisfies the definition of a canonical basis of ``e'' type.

\vspace{12.pt}

Further, according to \eqref{h}, the``h'' type
bases look as
\begin{equation}\nu_{N[N_1,N_2]}^n(\xi_1,\eta_1;\xi_2,\eta_2)=\hat{\rho}(N,N_1,N_2)(\xi_1\eta_2-\eta_1\xi_2)^{N_1+N_2-N}\cdot\end{equation}$$\cdot
\left(-\xi_2\frac{\partial}{\partial\xi_1}-\eta_2\frac{\partial}
{\partial\eta_1}\right)^{N-N_1+N_2}h_N^n(\xi_1,\eta_1).$$ These
bases are related as
\begin{equation}\label{mu_nu}\mu_{N[N_1,N_2]}^{\cdot}=
\nu_{N[N_1,N_2]}^{\cdot}V_N.\end{equation}

The Clebsch-Gordan matrices  $C_{N[N_1,N_2]}^n$ and
$G_{N[N_1,N_2]}^{n}$ are introduced as the coefficients of
decomposition of the basis polynomials $\mu_{N[N_1,N_2]}^n$ and
$\nu_{N[N_1,N_2]}^n$ by the polynomials
$\{e_{N_1}^{n_1}e_{N_2}^{n_2}\}$ and
$\{h_{N_1}^{n_1}h_{N_2}^{n_2}\}$, respectively:
\begin{equation}\label{C}\mu_{N[N_1,N_2]}^n=e_{N_1}^{\cdot}
C_{N[N_1,N_2]}^n(e_{N_2}^{\cdot})^T,\end{equation}
\begin{equation}\label{G}\nu_{N[N_1,N_2]}^n=h_{N_1}^{\cdot}
G_{N[N_1,N_2]}^n(h_{N_2}^{\cdot})^T\end{equation}

It is easy to see \cite{gord2}, that the matrices $C_{N[N_1,N_2]}^n,\
n=-N,\ldots,N$ and $G_{N[N_1,N_2]}^n$, $\ n=-N,\ldots,N$ constitute
canonical bases of type ``e'' and of type ``h'', respectively, in
the subspace of $(2N_1+1)\times(2N_2+1)$-dimensional matrices,
where an irreducible representation of dimension $N,\
N=|N_1-N_2|,\ldots,N_1+N_2$ is acting.

The following formula \cite{gord2}, which is a corollary of the
formulas \eqref{mu}, \eqref{C}, will be useful for computations of
the nonzero Clebsch-Gordan coefficients in the next section. For
all $n\in\{-N,-N+1,\ldots,N\}$, $n_2\in\{-N_2,-N_2+1,\ldots,N_2\}$ we have
$$\left\{\rho(N,N_1,N_2)\left(\frac{\partial}{\partial\xi_0}
\frac{\partial}{\partial\eta}-
\frac{\partial}{\partial\eta_0}\frac{\partial}{\partial\xi}
\right)^{N-N_1+N_2}e_N^n(\xi_0,\eta_0)\right\}\mid_{\xi_0=\xi,\eta_0=\eta}
\cdot$$\begin{equation}\cdot(-1)^{N_2-n_2}e_{N_2}^{-n_2}(\xi,\eta)
\label{c_comp}=\sum\limits_{n_1}
c_{N[N_1,N_2]}^{n[n_1,n_2]}e_{N_1}^{n_1}(\xi,\eta),\end{equation} where
\begin{equation}\label{rho}
\rho(N,N_1,N_2)=\sqrt{\frac{(2N_1+1)!(N_1+N_2-N)!(N+N_1-N_2)!}
{(2N)!(2N_2+1)!(N+N_1+N_2+1)!(N-N_1+N_2)!}}.\end{equation}

\section{Computing the Clebsch-Gordan matrices}
We want to compute the nonzero elements of the matrices
\begin{equation}G_{N[N_1,N_2]}^{\pm n},\ n=0,\ldots,N,\
N=|N_1-N_2|,|N_1-N_2|+1,\ldots,N_1+N_2\end{equation} for arbitrary integer
weights $N_1,N_2$. Note that in a similar way we can simply obtain
the matrices $C_{N[N_1,N_2]}^{\pm n}$.

The general scheme of computation, which we present in more
details below, is as follows:

 1) Compute the nonzero diagonal of the
two matrices $C_{N[N_1,N_2]}^{\pm N},$ by means of \eqref{c_comp}.

2) Compute the two matrices $G_{N[N_1,N_2]}^{\pm N}$ by means of
the transformation formulas \eqref{e_h} and \eqref{mu_nu}.

3) Compute other matrices by means of recurrent formulas
\eqref{h_down}, \eqref{h_up}, which allow  to define
$G_{N[N_1,N_2]}^{\pm (n-1)}$ from $G_{N[N_1,N_2]}^{\pm n}$.

\vspace{12.pt}

Now let us describe these three stages more precisely.

\vspace{8.pt}

$\bullet$ To carry out the first stage  we use the well-known
property of the Clebsch-Gordan coefficients stating that
\begin{equation}c_{N[N_1,N_2]}^{n[n_1,n_2]}=0\ \text{for}\ n_1+n_2\neq n,\end{equation} i.e.
each matrix $C_{N[N_1,N_2]}^n$ possesses only one nonzero
diagonal, thus the sum in the right-hand part of \eqref{c_comp},
for $n=N$, consists of only one nonzero summand (when $n_1=n-n_2$).

More precisely, for $n=N$ the elements of the nonzero diagonal,
have the numbers $(N-k,k)$, where $k=N-N_1,N-N_1+1,\ldots,N_2.$
Rewriting \eqref{c_comp} and taking into account \eqref{rho}
and \eqref{e}, we obtain
\begin{equation}c_{N[N_1,N_2]}^{N[N-k,k]}=(-1)^{N_2-k}\rho(N,N_1,N_2)
\frac{\tilde{\rho}(N_2,-k)}{\tilde{\rho}(N_1,N-k)}\sqrt{2N+1}\cdot\end{equation}
$$\cdot[(N+N_1-N_2+1)\ldots(2N-1)(2N)]\cdot[(N_1-N+k+1)
\ldots(N_2+k-1)(N_2+k)].$$

For $n=-N$, the elements of the nonzero diagonal have numbers
$(-N-k,k),\ \ k=-N_2,-N_2+1,\ldots,N_1-N$, and we obtain
\begin{equation}c_{N[N_1,N_2]}^{-N[-N-k,k]}=(-1)^{N_2-k}\rho(N,N_1,N_2)
\frac{\tilde{\rho}(N_2,-k)}{\tilde{\rho}(N_1,-N-k)}\sqrt{2N+1}\cdot\end{equation}
$$\cdot[(N+N_1-N_2+1)\ldots(2N-1)(2N)]\cdot[(N_1-N-k+1)
\ldots(N_2-k-1)(N_2-k)].$$

The case $N=0$ (and $N_2=N_1$) is considered
independently:
\begin{equation}c_{0[N_1,N_1]}^{0[-k,k]}=\frac{(-1)^{N_1-k}}{\sqrt{2N_1+1}},\
k=-N_1,-N_1+1,\ldots,N_1.\end{equation}

\vspace{12.pt}

$\bullet$ Now turn to the second stage.
From the formulas \eqref{mu_nu} and \eqref{C} we have
$$\nu_{N[N_1,N_2]}^N=\frac{-(-i)^N}{\sqrt{2}}[(-1)^N
\mu_{N[N_1,N_2]}^N+\mu_{N[N_1,N_2]}^{-N}]=\vspace{8.pt}$$
$$=\frac{-(-i)^N}{\sqrt2}[(-1)^Ne_{N_1}^{\cdot}C_{N[N_1,N_2]}^N(e_{N_2}^{\cdot})^T+e_{N_1}^{\cdot}C_{N[N_1,N_2]}^{-N}(e_{N_2}^{\cdot})^T]=\vspace{8.pt}$$
$$=\frac{-(-i)^N}{\sqrt2}h_{N_1}^{\cdot}V_{N_1}\ [(-1)^NC_{N[N_1,N_2]}+C_{N[N_1,N_2]}^{-N}]\ (V_{N_2})^T(h_{N_2}^{\cdot})^T,\vspace{8.pt}$$
hence \begin{equation}G_{N[N_1,N_2]}^N=\frac{-(-i)^N}{\sqrt{2}}V_{N_1}
\left\{(-1)^NC_{N[N_1,N_2]}^N+C_{N[N_1,N_2]}^{-N}\right\}(V_{N_2})^T.\end{equation}
In a similar way,
\begin{equation}G_{N[N_1,N_2]}^{-N}=\frac{(-i)^{N-1}}{\sqrt{2}}V_{N_1}
\left\{(-1)^NC_{N[N_1,N_2]}^N-C_{N[N_1,N_2]}^{-N}\right\}(V_{N_2})^T.\end{equation}
For $N=0\ (N_2=N_1)$, we have
\begin{equation}G_{0[N_1,N_1]}^{0}=V_{N_1}C_{0[N_1,N_1]}^0(V_{N_1})^T.\end{equation}

Note that the matrices calculated on this stage are real-valued.

\vspace{12.pt}

$\bullet$ Finally, in order
to compute other matrices $G_{N[N_1,N2]}^{\pm n},\ n<N$, on the third stage of the algorithm, we need
to compute first the action of the infinitesimal operators
$J_{\pm1}$, see \eqref{inf_oper}, \eqref{h_down}, \eqref{h_up}.

By definition,
\begin{equation}J_n=\frac{d}{d\omega}T_{g_n(\omega)}|_{\omega=0},\ \ n=-1,0,1,\end{equation}
where $g_n(\omega)$ are one-parameter subgroups of $SU(2)$.

As already mentioned in the previous section, in the space of spinor polynomials, $J_{\pm 1}$ are given by
formulas
\begin{equation}J_{-1}=\frac{1}{2}\left(-\eta\frac{\partial}{\partial\xi}+\xi
\frac{\partial}{\partial\eta}\right),\
J_1=\frac{i}{2}\left(\eta\frac{\partial}{\partial\xi}+
\xi\frac{\partial}{\partial\eta}\right)\end{equation}

From particular cases of the formula \eqref{c_comp}, written in
the basis ``h'', it follows that
\begin{equation}\frac{(-1)^n\sqrt{3}J_n}{\sqrt{N(N+1)(2N+1)}}h_N^k(\xi,\eta)=
\sum\limits_jg_{1[N,N]}^{n[j,k]}h_N^j(\xi,\eta),\end{equation} and hence
\begin{equation}\label{inf}J_{\pm1}^N=-\sqrt{\frac{N(N+1)(2N+1)}{3}}
G_{1[N,N]}^{\pm 1},
\end{equation}
where by $J_{\pm1}^N$ we mean the corresponding infinitesimal
operators of the representations of weight $N$. This formulas are
constructive, since the matrices $G_{1[N,N]}^{\pm 1}$ were
computed on the first stage.

Let us again consider the space of $(2N_1+1)\times(2N_2+1)$
matrices, which is isomorphic to the space of bispinor polynomials
and denote there the infinitesimal operators as
$J_{\pm1}^{N_1\times N_2}$. Recall that in this space the
Clebsch-Gordan matrices
\begin{equation}G_{N[N_1,N_2]}^{\pm n},\ n=0,\ldots,N,\
N=|N_1-N_2|,|N_1-N_2|+1,\ldots,N_1+N_2\end{equation} constitute a canonical
basis of ``h'' type.

The action of the operator $J_{\pm1}^{N_1\times N_2}$ on the
matrix $B$ of dimension $(2N_1+1)\times(2N_2+1)$ is computed as
$$J_{\pm1}^{N_1\times N_2}B=
-\frac{1}{\sqrt{3}}\{\sqrt{N_1(N_1+1)(2N_1+1)}G_{1[N_1,N_1]}^{\pm1}B+$$
\begin{equation}\sqrt{N_2(N_2+1)(2N_2+1)}B(G_{1[N_2,N_2]}^{\pm1})^T\}.\end{equation}

\vspace{8.pt}

It follows from the definition of the Kronecker product and the
infinitesimal operator and taking into account the formula
\eqref{inf}:
\begin{equation}(T_g^1\times T_g^2)B=T_g^1B(T_g^2)^T,\end{equation}
\begin{equation}J_{\pm1}^{N_1\times N_2}B=\frac{d}{d\omega}(T^1_{g_{\pm1}(\omega)}\times
 T^1_{g_{\pm1}(\omega)})|_{\omega=0}B=
J_{\pm1}^{N_1}B+BJ_{\pm1}^{N_2}.\end{equation}

Finally, the recurrent formulas \eqref{h_down} and \eqref{h_up}
allow to compute the matrices $G_{N[N_1,N_2]}^{\pm (n-1)}$ from
$G_{N[N_1,N_2]}^{\pm n}$.

 More precisely, for $N\geq n\geq 2$ we have
$$G_{N[N_1,N_2]}^{-(n-1)}=-\frac{1}{\sqrt{(N+n)(N-n+1)}}
\{J_{-1}^{N_1\times N_2}G_{N[N_1,N_2]}^{-n}+ $$
\begin{equation}+J_{1}^{N_1\times
N_2}G_{N[N_1,N_2]}^{n}\},\end{equation}
$$G_{N[N_1,N_2]}^{(n-1)}=-\frac{1}{\sqrt{(N+n)(N-n+1)}}
\{-J_{1}^{N_1\times N_2}G_{N[N_1,N_2]}^{-n}+$$
\begin{equation}+J_{-1}^{N_1\times N_2}G_{N[N_1,N_2]}^{n}\}.\end{equation} For $n=1$ we
have
\begin{equation}G_{N[N_1,N_2]}^{0}=\frac{1}{\sqrt{2N(N+1)}}
\left\{-J_{1}^{N_1\times N_2}G_{N[N_1,N_2]}^{-1}+
J_{-1}^{N_1\times N_2}G_{N[N_1,N_2]}^{1}\right\}.\end{equation}

\vspace{12.pt}

Note also that we can  compute the Clebsch-Gordan matrices by this algorithm
without the spinor technique (by using only the matrix approach), computing at the
first stage  the eigenvectors of the infinitesimal
operator $J_0$ acting in the space of matrices.

\section{Applications to elasticity theory}

\subsection{Statement of the problem}

Let $T=||s_{ij}||$ (where $i,j=-1,0,1$) be a tensor related to the
elastic media, e.g. in \cite{ss08} we consider the Piola-Kirchhoff
stress tensor (which is not necessarily symmetric) in Lagrangian
coordinates $\xi_{-1},\xi_0,\xi_1$, but we also may consider the
symmetric Cauchy stress tensor 
in Eulerian coordinates $x_{-1},x_0,x_1$, or some relevant function
of the distorsion matrix
$C=||c_{ij}||=||\frac{\partial x_i}{\partial\xi_j}||,$ whatever is more
convenient for a particular problem. Here we assume for simplicity
this tensor be symmetric: $T=T^*$, and use, when considering different crystal systems, a technique a bit
different from the one used in \cite{ss08}.

The internal energy  of the considered media can then be written
as a function of $\operatorname{det}(T)$ and five quadratic
invariants, which will be constructed below:
\begin{equation}\label{H}H=H(\operatorname{det}(T),J_0,J_1,I_0,I_1,I_2).\end{equation}
For linear elasticity theory, the internal energy is constructed just as the sum of the quadratic invariants.

The symmetric tensor $T$ can be written as
\begin{equation}\label{T} T=\left[\begin{array}{ccc}
s_{11}&s_{12}&s_{13}\\
s_{21}&s_{22}&s_{23}\\
s_{31}&s_{32}&s_{33}
\end{array}\right]= pI_3+S.
\end{equation}
 Here $p$
 is the ``pressure'', a
scalar, i.e. it is transformed by an irreducible representation of
weight 0;
\newline $S=S^*,\ tr(S)=0$ is the ``deviator'' matrix, consisting of five independent
elements, which form a vector transformed by an irreducible
representation of weight 2;
\newline $I_3$ is the identity matrix.

The matrix $S$ can be rewritten by means of the corresponding Clebsch-Gordan
matrices (which are a canonical basis in the space of symmetric $3\times 3$
matrices with zero trace) as follows:
$$S=\left[\begin{array}{ccc}
s_{11}-p&\frac{s_{12}+s_{21}}{2}&\frac{s_{13}+s_{31}}{2}\\
\frac{s_{12}+s_{21}}{2}&s_{22}-p&\frac{s_{23}+s_{32}}{2}\\
\frac{s_{13}+s_{31}}{2}&\frac{s_{23}+s_{32}}{2}&s_{33}-p\\
\end{array}\right] =$$
\begin{equation}=s_{-2}G_{2[1,1]}^{-2}+s_{-1}G_{2[1,1]}^{-1}
+s_{0}G_{2[1,1]}^{0}+s_{1}G_{2[1,1]}^{1}+ s_{2}G_{2[1,1]}^{2}.\end{equation}
Let us denote
 \begin{equation}\label{ves2}{\bf
s}=\left(\begin{array}{c}s_{-2}\\
s_{-1}\\
s_{0}\\
s_{1}\\
s_{2}\end{array}\right)=\left(\begin{array}{c}
-\displaystyle\frac{s_{13}+s_{31}}{\sqrt{2}}\vspace{6.pt}\\
\displaystyle\frac{s_{12}+s_{21}}{\sqrt{2}}\vspace{6.pt}\\
\displaystyle\frac{\sqrt{3}}{\sqrt{2}}(s_{22}-p)\vspace{6.pt}\\
\displaystyle\frac{s_{23}+s_{32}}{\sqrt{2}}\vspace{6.pt}\\
\displaystyle\frac{s_{11}-s_{33}}{\sqrt{2}}
\end{array}\right).\end{equation} This vector is transformed by 
weight 2, and the corresponding Clebsch-Gordan matrices (calculated
by the algorithm from the previous subsection) look as follows:
$$ G_{2[1,1]}^{-2}=\left[\begin{array}{ccc}
 0     &    0       &  -\frac{1}{\sqrt{2}}  \\
         0 &   0   &      0  \\
         -\frac{1}{\sqrt{2}}  &       0&   0
 \end{array}\right],\
 G_{2[1,1]}^{-1}=\left[\begin{array}{ccc}
 0     &    \frac{1}{\sqrt{2}}       &  0  \\
         \frac{1}{\sqrt{2}} &   0   &      0  \\
         0  &       0&   0
\end{array}\right],$$
$$G_{2[1,1]}^{0}=\left[\begin{array}{ccc}
-\frac{1}{\sqrt{6}}     &    0       &  0  \\
         0 &   \frac{2}{\sqrt{6}}  &      0  \\
         0  &       0&   -\frac{1}{\sqrt{6}}
\end{array}\right],$$
$$G_{2[1,1]}^{1}=\left[\begin{array}{ccc}
 0     &    0       &  0  \\
         0 &   0   &      \frac{1}{\sqrt{2}}  \\
         0  &       \frac{1}{\sqrt{2}}&   0
 \end{array}\right],\
 G_{2[1,1]}^{2}=\left[\begin{array}{ccc}
 \frac{1}{\sqrt{2}}     &    0       &  0  \\
         0 &   0   &      0  \\
         0  &       0&   -\frac{1}{\sqrt{2}}
\end{array}\right].$$

Let us calculate the matrix of the irreducible representation of
$SO(3)$ acting in $\mathbb R^5$ as
\begin{equation}\tilde{{\bf s}}=A{\bf s}\end{equation}
which transforms the above introduced  vector
 $${\bf s}=\left(s_{-2},s_{-1},s_{0},s_{1},s_{2}\right)^T$$
 (cf. \cite{gord1,gelf,vil} etc. where the matrix entries of
 irreducible are calculated with different
 methods for the general case).
The matrix $A$ can be represented as the product
$A_{\theta}A_{\varphi}A_{\psi}$ of the matrices of representations
of the rotations $U_{\theta}$,$U_{\varphi}$, $U_{\psi}$ around the
coordinate axes (as it is well-known, any element $U$ of $SO(3)$ is a
combination of such rotations).

The transformation  $${\bf x}^{\prime}=U{\bf x}$$ of the space of
vectors $${\bf x}=(x_{-1},x_0,x_1)^T$$ induces the transformation
$U^TSU$ of the tensor $S$, thus in order to calculate the elements
$A_{ij}$ of the desired matrix  $A$ we first calculate
independently each matrix
$U^TG^j_{2[1,1]}U$ of the sum
\begin{equation}\label{s}\tilde{S}=\sum\limits_{j=-2}^2s_jU^TG^j_{2[1,1]}U=\sum\limits_{i=-2}^2\tilde{s}_i
G^i_{2[1,1]}\end{equation} and  deduce from \eqref{s}
\begin{equation}\label{s}\tilde{s}_i=\tilde{s}_i(s_j,\varphi,\psi,\theta)=
\sum\limits_{i=-2}^2A_{ij}s_j.\end{equation} For the first one-parameter subgroup
\begin{equation}
U=U_{\theta}= \left[\begin{array}{ccc}
\cos(\theta) & -\sin(\theta) & 0\\
\sin(\theta) & \cos(\theta)  & 0\\
0            & 0             & 1
\end{array}\right]
\end{equation}
(around the axis $x_{-1}$)  we obtain
 the
following matrix
\small{\begin{equation}\label{A}
A_{\theta}= \left[\begin{array}{ccccc}
\cos(\theta) & 0 & 0  & -\sin(\theta) & 0 \\
0 & \cos(2\theta) & \frac12\sqrt3\sin(2\theta) & 0 & -\frac12\sin(2\theta) \\
0 & -\frac12\sqrt3\sin(2\theta) & \cos(\theta)^2 -
\frac12\sin(\theta)^2 & 0
& \frac12\sqrt3\sin(\theta)^2 \\
\sin(\theta) & 0 & 0  & \cos(\theta) & 0 \\
0 & \frac12\sin(2\theta) & \frac12\sqrt3\sin(\theta)^2 & 0 &
\frac12 + \frac12\cos(\theta)^2
\end{array}\right].
\end{equation}}

The two other subgroups are treated in
a similar way. The knowledge of the obtained matrices will be helpful when considering different crystal systems below.

\vspace{12.pt}

\subsection{Construction of invariants}

To write down all of the invariants of the variables $\{s_{ij}\}$
in \eqref{H}, let us first ask, which quadratic forms (and how
many) can be constructed from the elements $p$ and $\bf{s}$ from
\eqref{ves2}. For this purpose we write all possible Kronecker
products of the weights 0 and 2
$$0\times 0,\ 0\times 2,\ 2\times 0,\ 2\times 2$$ and their decompositions into irreducible
ones. Note that, due to commutativity of the Kronecker product,
the cases $0\times 2$ and $2\times 0$
 are identical. According to \eqref{vesa}, from the decomposition of the product of weights
  $N_1$ and $N_2$ into irreducible ones
\begin{equation}\label{sum}\left[ 2|N_1-N_2|+1\right]+\ldots
+\left[2(N_1+N_2)+1\right]=(2N_1+1)(2N_2+1)\end{equation}
parameters ${\bf w}_n^{(N)}$ arise, which transform by irreducible
representations of the corresponding weights $N$.
This fact is illustrated by the following formula
\cite{gg}:
\begin{equation}\label{kronek}{\bf p}^{(N_1)}\times{\bf
q}^{(N_2)}=\sum\limits_{N=|N_1-N_2|}^{N_1+N_2}\left(\sum\limits_{n=-N}^N
{\bf w}_n^{(N)}G_{N[N_1,N_2]}^n\right),\end{equation} where the
vectors ${\bf p}$ and ${\bf q}$ are transformed by means of
irreducible representations of weights $N_1$ and $N_2$,
respectively.

 The invariant quadratic forms (transformed by a
representation of zero weight) made from these vectors look as
follows (their invariance is an obvious consequence of 
orthogonality of the representations under consideration):
\begin{equation}\label{inv} I_{(N)}=\sum\limits_{n=-N}^N{\bf w}^{(N)}_n\left(\left[{\bf
p}^{(N_1)}\right]^TG_{N[N_1,N_2]}^{n},{\bf
q}^{(N_2)}\right),\end{equation}
$$N=|N_1-N_2|,\ |N_1-N_2|+1,\ldots,N_1+N_2.$$
 In this way, we can calculate how many parameters (the coefficients of these quadratic
 forms) characterizing the elastic media there are and write down
 all possible invariant quadratic forms in \eqref{H}. During this
 process we will identify the coinciding (due to commutativity)
 invariants and remove the skew-symmetric ones (for the odd sum of weights $N+N_1+N_2$).
 Finally we will obtain, according to
\eqref{sum},

$1)\ 0\times 0\Longrightarrow 1$ parameter;

$2)\ 0\times 2, 2\times 0\Longrightarrow 5$ parameters;

$3)\ 2\times 2\Longrightarrow 1+5+9=15$ parameters (for the
weights $0,2,4$; note that the cases of weights $1,3$ are not
meaningful because of the skew-symmetry of the korresponding
Clebsch-Gordan matrices),

thus 21  parameters characterizing the elastic media, which is in
accordance with the classical elasticity theory.

According to \eqref{inv}, the corresponding invariant quadratic
forms look as follows.

\vspace{8.pt}

 1) For $ N1=0;\ N2=0;\ N=0$ we have
\begin{equation}J_{0}=c_1G_{0[0,0]}^0p^2=c_1p^2.\end{equation}

\noindent 2) For $ N1=0;\ N2=2;\ N=2$ the Clebsch-Gordan matrices
look again very simple: $G_{2[0,2]}^{0}=\left[ 0\ 0\ 1\ 0\
0\right],\ G_{2[0,2]}^{-1}=\left[ 0\ 1\ 0\ 0\ 0\right],\
G_{2[0,2]}^{1}=\left[ 0\ 0\ 0\ 1\ 0\right],\
G_{2[0,2]}^{-2}=\left[ 1\ 0\ 0\ 0\ 0\right],\
G_{2[0,2]}^{2}=\left[ 0\ 0\ 0\ 0\ 1\right],$ and the correspondinq
quadratic form has five arbitrary parameters $a_j,
j=-2,-1,\ldots,2$:
\begin{equation}\label{J1}J_1=\sum\limits_{j=-2}^{2}a_jG_{2[0,2]}^j{\bf
s}p =\sum\limits_{j=-2}^{2}a_js_jp.\end{equation} The same expression appears
from the case $ N1=2;\ N2=0;\ N=2$.

\noindent 3) For the case $ N1=2;\ N2=2$, as explained above, we
have to consider $N=0,2,4$.

$\bullet$ For $\ N=0 $ we have \begin{equation}I_{0}=c_2(G_{0[2,2]}^0\ {\bf s},
{\bf s})=\tilde{c}_2({\bf s}, {\bf s}),\end{equation} where $G_{0[2,2]}=cI_3$
is a diagonal matrix.

$\bullet$ For $ N1=2;\ N2=2;\ N=2$ the invariant quadratic form
looks like \begin{equation}\label{I1}I_{1}=\sum\limits_{j=-2}^{2}b_j(G_{2[2,2]}^j {\bf s},
{\bf s}),\end{equation} where
$$ G_{2[2,2]}^{-2}=\left[\begin{array}{ccccc}
       0&       0&   -\frac{\sqrt{2}}{\sqrt{7}}&       0&         0\\
       0&       0&       0&   -\frac{\sqrt{3}}{\sqrt{2}\sqrt{7}}&         0\\
   -\frac{\sqrt{2}}{\sqrt{7}}&       0&       0&       0&         0\\
       0&   -\frac{\sqrt{3}}{\sqrt{2}\sqrt{7}}&       0&       0&         0\\
       0&       0&       0&       0&         0
\end{array}\right],$$ $$
 G_{2[2,2]}^{-1}=\left[\begin{array}{ccccc}
       0&       0&       0&   -\frac{\sqrt{3}}{\sqrt{2}\sqrt{7}}&       0\\
       0&       0&    \frac{1}{\sqrt{2}\sqrt{7}}&       0&    \frac{\sqrt{3}}{\sqrt{2}\sqrt{7}}\\
       0&    \frac{1}{\sqrt{2}\sqrt{7}}&       0&       0&       0\\
   -\frac{\sqrt{3}}{\sqrt{2}\sqrt{7}}&       0&       0&       0&       0\\
       0&    \frac{\sqrt{3}}{\sqrt{2}\sqrt{7}}&       0&       0&       0
\end{array}\right],$$
$$G_{2[2,2]}^{0}=\left[\begin{array}{ccccc}
 -\frac{\sqrt{2}}{\sqrt{7}}&       0&     0&       0&       0\\
     0&    \frac{1}{\sqrt{2}\sqrt{7}}&     0&       0&       0\\
     0&       0&  \frac{\sqrt{2}}{\sqrt{7}}&       0&       0\\
     0&       0&     0&    \frac{1}{\sqrt{2}\sqrt{7}}&       0\\
     0&       0&     0&       0&   -\frac{\sqrt{2}}{\sqrt{7}}
\end{array}\right],$$
$$
G_{2[2,2]}^{1}=\left[\begin{array}{ccccc}
       0&   -\frac{\sqrt{3}}{\sqrt{2}\sqrt{7}}&       0&       0&       0\\
   -\frac{\sqrt{3}}{\sqrt{2}\sqrt{7}}&       0&       0&       0&       0\\
       0&       0&       0&    \frac{1}{\sqrt{2}\sqrt{7}}&       0\\
       0&       0&    \frac{1}{\sqrt{2}\sqrt{7}}&       0&   -\frac{\sqrt{3}}{\sqrt{2}\sqrt{7}}\\
       0&       0&       0&   -\frac{\sqrt{3}}{\sqrt{2}\sqrt{7}}&       0
\end{array}\right],$$  $$G_{2[2,2]}^{2}=\left[\begin{array}{ccccc}
    0&       0&       0&       0&       0\\
    0&    \frac{\sqrt{3}}{\sqrt{2}\sqrt{7}}&       0&       0&       0\\
    0&       0&       0&       0&   -\frac{\sqrt{2}}{\sqrt{7}}\\
    0&       0&       0&   -\frac{\sqrt{3}}{\sqrt{2}\sqrt{7}}&       0\\
   0&       0&   -\frac{\sqrt{2}}{\sqrt{7}}&       0&       0
\end{array}\right].
$$

$\bullet$ Finally, for the case $ N1=2;\ N2=2;\ N=4 $ we obtain
 \begin{equation}\label{I2}I_{2}=\sum\limits_{j=-4}^{4}d_j(G_{4[2,2]}^j
{\bf s}, {\bf s}), \end{equation} where
$$G_{4[2,2]}^{-4}=\left[\begin{array}{ccccc}
       0&       0&   0&       0&         -\frac{\sqrt{2}}{2}\\
       0&       0&       0&   0&         0\\
   0&       0&       0&       0&         0\\
       0&   0&       0&       0&         0\\
      - \frac{\sqrt{2}}{2}&       0&       0&       0&         0
\end{array}\right],\quad G_{4[2,2]}^{-3}=\left[\begin{array}{ccccc}
       0&       0&   0&       -\frac{1}{2}&         0\\
       0&       0&       0&   0&         -\frac{1}{2}\\
  0&       0&       0&       0&         0\\
       -\frac{1}{2}&   0&       0&       0&         0\\
      0&        -\frac{1}{2}&       0&       0&         0
\end{array}\right],$$
$$ G_{4[2,2]}^{-2}=\left[\begin{array}{ccccc}
       0&       0&   \frac{\sqrt{14}\sqrt{3}}{14}&       0&         0\\
       0&       0&       0&   -\frac{\sqrt{14}}{7}&         0\\
  \frac{\sqrt{14}\sqrt{3}}{14}&       0&       0&       0&         0\\
       0&   -\frac{\sqrt{14}}{7}&       0&       0&         0\\
       0&       0&       0&       0&         0
\end{array}\right],$$ $$G_{4[2,2]}^{-1}=\left[\begin{array}{ccccc}
       0&       0&   0&       \frac{\sqrt{7}}{14}&         0\\
       0&       0&       \frac{\sqrt{7}\sqrt{3}}{7}&   0&         -\frac{\sqrt{7}}{14}\\
  0&       \frac{\sqrt{7}\sqrt{3}}{7}&       0&       0&         0\\
        \frac{\sqrt{7}}{14}&   0&       0&       0&         0\\
       0&       - \frac{\sqrt{7}}{14}&       0&       0&         0
\end{array}\right],$$
$$G_{4[2,2]}^{0}=\left[\begin{array}{ccccc}
       \frac{\sqrt{5}\sqrt{2}\sqrt{7}}{70}&       0&   0&       0&         0\\
       0&        -\frac{2\sqrt{5}\sqrt{2}\sqrt{7}}{35}&       0&   0&         0\\
   0&       0&       \frac{3\sqrt{5}\sqrt{2}\sqrt{7}}{35}&       0&         0\\
       0&   0&       0&       -\frac{2\sqrt{5}\sqrt{2}\sqrt{7}}{35}&         0\\
      0&       0&       0&       0&         \frac{\sqrt{5}\sqrt{2}\sqrt{7}}{70}
\end{array}\right],$$
$$\hspace{-40.pt}G_{4[2,2]}^{1}=\left[\begin{array}{ccccc}
       0&        \frac{\sqrt{7}}{14}&   0&      0&         0\\
        \frac{\sqrt{7}}{14}&       0&       0&   0&         0\\
  0&      0&       0&        \frac{\sqrt{7}\sqrt{3}}{7}&         0\\
        0&   0&       \frac{\sqrt{7}\sqrt{3}}{7}&       0&         \frac{\sqrt{7}}{14}\\
       0&       0&       0&       \frac{\sqrt{7}}{14}&         0
\end{array}\right],$$ $$ G_{4[2,2]}^{2}=\left[\begin{array}{ccccc}
       0&       0&   0&       0&         0\\
       0&       \frac{\sqrt{14}}{7}&       0&   0&         0\\
  0&       0&       0&       0&         \frac{\sqrt{14}\sqrt{3}}{14}\\
       0&   0&       0&       -\frac{\sqrt{14}}{7}&         0\\
       0&       0&       \frac{\sqrt{14}\sqrt{3}}{14}&       0&         0
\end{array}\right],$$
$$G_{4[2,2]}^{-3}=\left[\begin{array}{ccccc}
       0&       \frac{1}{2}&   0&      0&         0\\
       \frac{1}{2}&       0&       0&   0&        0\\
  0&       0&       0&       0&         0\\
      0&   0&       0&       0&         -\frac{1}{2}\\
      0&       0&       0&       -\frac{1}{2}&         0
\end{array}\right],\quad G_{4[2,2]}^{-4}=\left[\begin{array}{ccccc}
       \frac{\sqrt{2}}{2}&       0&   0&       0&         0\\
       0&       0&       0&   0&         0\\
   0&       0&       0&       0&         0\\
       0&   0&       0&       0&         0\\
      0&       0&       0&       0&         -\frac{\sqrt{2}}{2}
\end{array}\right].$$

\vspace{12.pt}

As we can see, the 21 parameters characterizing the elastic media
have split into several groups:
\begin{itemize}
\item two scalars $c_1,\ c_2$;
\item two 5-dimensional vectors ${\bf a}=(a_{-2},a_{-1},a_0,a_1,a_2)$ and ${\bf b}=(b_{-2},b_{-1},b_0,b_1,b_2)$;
\item one 9-dimensional vector ${\bf d}=(d_{-4},d_{-3},d_{-2},d_{-1},d_0,d_1,d_2,d_3,d_4)$.
\end{itemize}

\subsection{Examples}

For the crystal systems, where there are additional symmetries,
the number of independent parameters is less than 21. Using again
the representation group approach, this fact can be established
easily for all of the seven crystal systems, also called syngonies. Below this is shown on the example of the monoclinic and rhombic
crystal systems.

Recall that a crystal group $\Delta$ consists of several
orthogonal transformations:
\begin{equation}\Delta\subset\{h:\ x\mapsto \tau+Bx\ |\ \tau\in\mathbb R^3,\ B\in
O(3)\},\end{equation} and $K$ is a {\it crystal} with the symmetry group
$\Delta$, if
\begin{equation}K=hK,\ \forall h\in\Delta.\end{equation}
Denote as $\Gamma$  the {\it group of rotations} of the crystal,
i.e. the set of matrices constituting the elements of $\Delta.$

For the triclinic system, when $\Gamma={I}$, we have the full set
of 21 independent parameters. For the monoclinic system we have
$\Gamma=<R_2>$, where \begin{equation}R_2=\left[\begin{array}{ccc}
-1  & 0 & 0 \\
0           & -1 & 0 \\
0& 0 & 1
\end{array}\right]=U_{\theta}\end{equation} is the rotation around the axis $x_{-1}$ on
$\theta=\pi$. In this case the transformation matrix \eqref{A}
looks like
\begin{equation}A=A_{\theta}=\left[\begin{array}{ccccc}
-1 & 0          & 0 & 0           & 0 \\
0           & 1 & 0 & 0 & 0            \\
0           & 0          & 1 & 0           & 0            \\
0           & 0 & 0 &-1  & 0            \\
0 & 0          & 0 & 0           & 1
\end{array}\right].\end{equation}
Taking in a account the invariance of the quadratic forms
introduced above, we arrive at the following relations: for the invariant $J_1$ from \eqref{J1},
$$\sum\limits_{j=-2}^2a_js_j=a_{-2}s_{-2}+a_{-1}s_{-1}+a_0s_0+a_1s_1+a_2s_2=$$$$=-a_{-2}s_{-2}+a_{-1}s_{-1}+a_0s_0-a_1s_1+a_2s_2=\sum\limits_{j=-2}^2a_j\tilde{s}_j,$$
from where it follows that \begin{equation}a_{-2}=a_1=0.\end{equation} The identity for the invariant $I_1$ from \eqref{I1}, looking as
$$\sum\limits_{j=-2}^2b_j<G^j_{2[2,2]}{\bf s},{\bf s}>=\sum\limits_{j=-2}^2b_j<G^j_{2[2,2]}
{\bf \tilde{s}}, {\bf
\tilde{s}}>=\sum\limits_{j=-2}^2b_j<A^TG^j_{2[2,2]}A{\bf s},{\bf
s}>,$$ gives \begin{equation}b_{-2}=b_{1}=0.\end{equation} Finally, the identity for the invariant $I_2$ from \eqref{I2} we have
$$\sum\limits_{j=-4}^4d_j<G^j_{4[2,2]}{\bf s},{\bf s}>=\sum\limits_{j=-4}^4d_j<G^j_{4[2,2]}
{\bf \tilde{s}}, {\bf
\tilde{s}}>=\sum\limits_{j=-4}^4d_j<A^TG^j_{4[2,2]}A{\bf s},{\bf
s}>,$$ from where
\begin{equation}d_{-4}=d_{-2}=d_1=d_3=0.\end{equation}
Consequently, for the monoclinic crystal system there are
$21-2-2-4=13$ independent parameters which characterize the media. The quadratic invariants look in this case as
$$J_0=c_1{\bf p}^2,\ J_1=\left(a_{-1}G_{2[0,2]}^{-1}+a_{0}G_{2[0,2]}^{0}+
a_{2}G_{2[0,2]}^{2}\right){\bf s}p,$$
 \begin{equation}I_{0}=c_2({\bf s},{\bf s}),\
I_{1}=\left<\left(b_{-1}G_{2[2,2]}^{-1}+b_{0}G_{2[2,2]}^{0}+b_{2}G_{2[2,2]}^{2}\right)
{\bf s},{\bf s}\right>,\end{equation}
$$I_{2}=\left<\left(d_{-3}G_{4[2,2]}^{-3}+d_{-1}G_{4[2,2]}^{-1}+d_{0}G_{4[2,2]}^{0}+
d_{2}G_{4[2,2]}^{2} +d_{4}G_{4[2,2]}^{4}\right) {\bf s},{\bf
s}\right>.$$

\vspace{8.pt}

For the rhombic system $<R_2,L_2>$, where
\begin{equation}L_2=\left[\begin{array}{ccc}
-1  & 0 & 0 \\
0           & 1 & 0 \\
0& 0 & -1\end{array}\right]=U_{\varphi}\end{equation} for $\varphi=\pi $,
we need one more matrix, corresponding to the rotation $L_2$:
\begin{equation}A=A_{\varphi}=\left[\begin{array}{ccccc}
1 & 0          & 0 & 0           & 0 \\
0           & -1 & 0 & 0 & 0            \\
0           & 0          & 1 & 0           & 0            \\
0           & 0 & 0 &-1  & 0            \\

0 & 0          & 0 & 0           & 1
\end{array}\right].\end{equation}
Besides the zero coefficients obtained for the monoclinic system above, we have
\begin{equation}a_{-1}=b_{-1}=d_{-1}=d_{-3}=0.\end{equation}

The invariants look now like $$J_0=c_1{\bf p}^2,\ J_1=a_{0}G_{2[0,2]}^{0}+a_2G_{2[0,2]}^{2}
{\bf s}p,$$
 \begin{equation}I_{0}=c_2({\bf s},{\bf s}),\
I_{1}=\left<\left(b_{0}G_{2[2,2]}^{0}+b_{2}G_{2[2,2]}^{2}\right)
{\bf s},{\bf s}\right>,\end{equation} $$I_{2}=\left<\left(d_{0}G_{4[2,2]}^{0}+
d_{2}G_{4[2,2]}^{2} +d_{4}G_{4[2,2]}^{4}\right) {\bf s},{\bf
s})\right>.$$

\vspace{8.pt}

For the five other crystal systems the independent parameters are
calculated in a similar way. By another method, using the contravariance of the Hook's tensor, this was done in \cite{ss08}.

For the isotropic case the only quadratic invariants are
$$J_0=c_1{\bf p}^2,\  I_{0}=c_2({\bf s},{\bf s}).$$

\vspace{8.pt}

Finally, let us consider the example of linear elasticity theory,
when the equations can be written as follows:
\begin{equation}\begin{cases}
\rho\displaystyle\frac{\partial u_i}{
\partial t}- \frac{\partial\sigma _{ij} } { \partial x_j}=0,\vspace{6.pt}\\
 \displaystyle\frac{\partial \varepsilon _{ij}}{ \partial t}-
 \frac{1}{2}(\frac{ \partial u_i } {\partial x_j}+
\frac{\partial u_j} {\partial x_i}) =0,
\end{cases}\end{equation}
where $i\ ,j=1\ ,\ 2,\ 3,$ $u_i$ are the velocities,
$\varepsilon _{ij}=\varepsilon _{ji}$ is the tensor of
deformations and $\sigma_{ij}=\sigma_{ji}$ is the tensor of stresses.
From the Hook's law $\varepsilon _{ij}=c^{ijkl}\sigma _{kl}$,
\begin{equation} \label{eq_0}
 \begin{cases}
  \displaystyle\rho\frac{\partial u_i}{
\partial t}- \frac{\partial\sigma _{ij} } { \partial x_i}=0,
    \vspace{6.pt}\\
\displaystyle c^{ijkl}\frac{\partial \sigma _{kl}}{ \partial t}-
 \frac{1}{2}(\frac{ \partial u_i } {\partial x_j}+
\frac{\partial u_j} {\partial x_i}) =0.
\end{cases}
\end{equation}
It is a symmetric hyperbolic system (in the sense of Friedrichs
\cite{fri}), which may written, by means of Clebsch-Gordan
matrices, in an invariant form (here ${\bf v}=(u_{-1},u_0,u_1)$ is
the velocity vector, $p$ is pressure, ${\bf s}$ is the 5-dimensional
vector defined in \eqref{ves2}):

\begin{equation}\label{eq_dec_an}
\begin{cases}
A_1\displaystyle\frac{\partial}{\partial t}{\bf v}+\Delta_{-}{\bf
s}+\Delta_{+}{\bf
p}=0,\\\vspace{12.pt}
\hat{A}\frac{\partial }{\partial
t}\left(\begin{array}{c}{\bf p}\\ \bf{s}\end{array}\right)+\left(\begin{array}{cc}\Delta_{-}{\bf v^{(1)}}&0\\0&
\Delta_{+}{\bf v^{(1)}}\end{array}\right)=0.\end{cases}
\end{equation}

Here 
$$\Delta_{-}{\bf
u^{(L)}}=c_{-}(L)\sum\limits_{i=-1}^1\displaystyle\frac{\partial}{\partial
x_i}G^i_{1[L-1,L]}{\bf u^{(L)}},\ \ \ \Delta_{+}{\bf
u^{(L)}}=c_{+}(L)\sum\limits_{i=-1}^1\displaystyle\frac{\partial}{\partial
x_i}G^i_{1[L+1,L]}{\bf u^{(L)}}$$ are invariant under rotations
matrix differential operators. The first of them lowers the weight
of the vector from $L$ to $L-1$ (an analog of  $div$), the second
one makes the weight bigger, from $L$ to $L+1$ (an analog of
$grad$).
The matrix $\hat{A}$ looks as
\small{\begin{equation}\left(\begin{array}{ccc}A_0&&0\\0&&A_2\end{array}\right)+\left(\begin{array}{ccc}0&
\sum\limits_{j=-2}^{2}a_{j}G_{2[0,2]}^j\\\sum\limits_{j=-2}^{2}a_{j}G_{2[2,0]}^j&\
\ \sum\limits_{j=-2}^{2}b_jG_{2[2,2]}^j
+\sum\limits_{j=-4}^{4}d_jG_{4[2,2]}^j\end{array}\right)\end{equation}}

Here
$$A_0=\hat{c}_1G^0_{0[0,0]}=\hat{c}_1=\frac{1}{3\lambda+2\mu},\
A_2=\hat{c}_2G^0_{0[2,2]}=\left(\begin{array}{ccccc}\frac{1}{\mu}&&&&\\
&\frac{1}{\mu}&&&\\&&\frac{1}{\mu}&&\\&&&\frac{1}{\mu}&\\&&&&\frac{1}{\mu}
\end{array}\right),$$
where $\lambda$ and $\mu$ are the Lame coefficients In the isotropic case the constants $a_i,b_i$ and $d_j$ are equal to zero. For the triclinic system all of the 21 constants are present, while for other crystal systems some of them are equal to zero or are linearly dependent, as described above. In the case of nonlinear elasticity the systems \eqref{eq_0} and \eqref{eq_dec_an} look more complicated: the matrix by $\frac{\partial }{\partial
t}$ is constructed from second partial derivatives of the potential $H$ from \eqref{H}, which has to be a convex function, and several other equations may be needed to add to the system, depending on the model \cite{ss08,god,gp,gmr}.

\vspace{12.pt}

{\bf Acknowledgements.} I thank  S.K. Godunov for suggesting me this proble\-matic and fruitful discussions and to  V.M. Gordienko for a consultation on his papers.

\end{document}